\documentclass[letter,longauth]{aa} % for the long lists of affiliations 
\usepackage{graphicx}
\usepackage{subfigure}
\usepackage{txfonts}
\usepackage{multirow}
\usepackage{longtable,lscape}
\usepackage{tablefootnote}
\usepackage{hyperref}
\graphicspath{ {Figures/} }
\newcommand{\mearth}{$M_\oplus$}
\newcommand{\rearth}{$R_\oplus$}

\begin{document} 

\defcitealias{2024A&A...682A.129M}{MG24}

   \title{Orbital obliquity of the young planet TOI-5398 b \\%through the Rossiter-McLaughlin effect 
   and the evolutionary history of the system
   \thanks{Table A.1 is only available in electronic form at the CDS via anonymous ftp to \url{https://cdsarc.u-strasbg.fr} (130.79.128.5) or via \url{http://cdsweb.u-strasbg.fr/cgi-bin/qcat?J/A+A/}.}\thanks{Based on observations made with the Italian \textit{Telescopio Nazionale Galileo} (TNG) operated by the \textit{Fundación Galileo Galilei} (FGG) of the \textit{Istituto Nazionale di Astrofisica} (INAF) at the \textit{Observatorio del Roque de los Muchachos} (La Palma, Canary Islands, Spain).}}

   \author{G. Mantovan
          \inst{\ref{inst1},\ref{inst2}} $^{\href{https://orcid.org/0000-0002-6871-6131}{\includegraphics[scale=0.5]{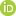}}}$ \and L. Malavolta\inst{\ref{inst1},\ref{inst2}}$^{\href{https://orcid.org/0000-0002-6492-2085}{\includegraphics[scale=0.5]{Figures/orcid.jpg}}}$ \and D. Locci\inst{\ref{inst3}}$^{\href{https://orcid.org/0000-0002-9824-2336}{\includegraphics[scale=0.5]{orcid.jpg}}}$ \and D. Polychroni\inst{\ref{inst4}}$^{\href{https://orcid.org/0000-0002-7657-7418}{\includegraphics[scale=0.5]{orcid.jpg}}}$ \and D. Turrini\inst{\ref{inst5}}$^{\href{https://orcid.org/0000-0002-1923-7740}{\includegraphics[scale=0.5]{orcid.jpg}}}$ \and A. Maggio\inst{\ref{inst3}}$^{\href{https://orcid.org/0000-0001-5154-6108}{\includegraphics[scale=0.5]{orcid.jpg}}}$ \and S. Desidera\inst{\ref{inst2}}$^{\href{https://orcid.org/0000-0001-8613-2589}{\includegraphics[scale=0.5]{orcid.jpg}}}$ \and R. Spinelli\inst{\ref{inst3}} \and S. Benatti\inst{\ref{inst3}}$^{\href{https://orcid.org/0000-0002-4638-3495}{\includegraphics[scale=0.5]{orcid.jpg}}}$\and G. Piotto\inst{\ref{inst1}}$^{\href{https://orcid.org/0000-0002-9937-6387}{\includegraphics[scale=0.5]{orcid.jpg}}}$ 
          \and A. F. Lanza\inst{\ref{inst15}}$^{\href{https://orcid.org/0000-0001-5928-7251}{\includegraphics[scale=0.5]{orcid.jpg}}}$\and F. Marzari\inst{\ref{inst1}} 
          \and A. Sozzetti\inst{\ref{inst5}}$^{\href{https://orcid.org/0000-0002-7504-365X}{\includegraphics[scale=0.5]{orcid.jpg}}}$
          \and M. Damasso\inst{\ref{inst5}}$^{\href{https://orcid.org/0000-0001-9984-4278}{\includegraphics[scale=0.5]{orcid.jpg}}}$ \and D. Nardiello\inst{\ref{inst1}}$^{\href{https://orcid.org/0000-0003-1149-3659}{\includegraphics[scale=0.5]{orcid.jpg}}}$
          \and L. Cabona\inst{\ref{inst24}}$^{\href{https://orcid.org/0000-0002-5130-4827}{\includegraphics[scale=0.5]{orcid.jpg}}}$\and M. D'Arpa\inst{\ref{inst3}}$^{\href{https://orcid.org/0009-0004-5914-7274}{\includegraphics[scale=0.5]{orcid.jpg}}}$ \and G. Guilluy\inst{\ref{inst5}}$^{\href{https://orcid.org/0000-0002-1259-2678}{\includegraphics[scale=0.5]{orcid.jpg}}}$  \and L. Mancini\inst{\ref{inst8}, \ref{inst9}, \ref{inst5}} \and G. Micela\inst{\ref{inst3}}$^{\href{https://orcid.org/0000-0002-9900-4751}{\includegraphics[scale=0.5]{orcid.jpg}}}$  \and V. Nascimbeni\inst{\ref{inst2}}$^{\href{https://orcid.org/0000-0001-9770-1214}{\includegraphics[scale=0.5]{orcid.jpg}}}$ \and T. Zingales\inst{\ref{inst1},\ref{inst2}}$^{\href{https://orcid.org/0000-0001-6880-5356}{\includegraphics[scale=0.5]{orcid.jpg}}}$
          }

    %TC:ignore
   \institute{Dipartimento di Fisica e Astronomia ``Galileo Galilei'', Università di Padova, Vicolo dell'Osservatorio 3, IT-35122, Padova, Italy\\
              \email{giacomo.mantovan@unipd.it}\label{inst1}
        \and INAF - Osservatorio Astronomico di Padova, Vicolo dell'Osservatorio 5, IT-35122, Padova, Italy\label{inst2}
        \and INAF - Osservatorio Astronomico di Palermo, Piazza del Parlamento 1, I-90134, Palermo, Italy\label{inst3}
        \and INAF - Osservatorio Astronomico di Trieste, Via Giambattista Tiepolo, 11, I-34131, Trieste (TS), Italy\label{inst4}
        \and INAF - Osservatorio Astrofisico di Torino, via Osservatorio 20, I-10025, Pino Torinese, Italy\label{inst5}
        \and INAF - Osservatorio Astrofisico di Catania, Oss. Astr. Catania, via S. Sofia 78, 95123 Catania Italy\label{inst15}
        \and INAF – Osservatorio Astronomico di Brera, Via E. Bianchi 46, 23807 Merate (LC), Italy\label{inst24}
        \and Department of Physics, University of Rome ``Tor Vergata'', Via della Ricerca Scientifica 1, I-00133, Rome, Italy\label{inst8}
        \and Max Planck Institute for Astronomy, K\"{o}nigstuhl 17, D-69117, Heidelberg, Germany\label{inst9}
        % \and Dipartimento di Scienza e Alta Tecnologia, Università degli Studi dell'Insubria, via Valleggio 11, 22100, Como, Italy\label{inst6}
        % \and INAF, Osservatorio Astronomico di Brera, Via E. Bianchi 46, 23807, Merate, Italy\label{inst7}
             }
             %TC:endignore

   \date{Compiled: \today}

% \abstract{}{}{}{}{} 
% 5 {} token are mandatory
 
  \abstract{Multi-planet systems exhibit remarkable architectural diversity. However, short-period giant planets are typically isolated. Compact systems like TOI-5398, with an outer close-orbit giant and an inner small-size planet, are rare among systems containing short-period giants. TOI-5398’s unusual architecture coupled with its young age (650 $\pm$ 150 Myr) make it a promising system for measuring the original obliquity between the orbital axis of the giant and the stellar spin axis in order to gain insight into its formation and orbital migration. %Unlike ordinary short-period giants, we can rule out the high-eccentricity migration scenario for compact systems such as TOI-5398.

  We collected in-transit (plus suitable off-transit) observations of TOI-5398 b with HARPS-N at TNG on March 25, 2023, obtaining high-precision radial velocity time series that allowed us to measure the Rossiter-McLaughlin (RM) effect. By modelling the RM effect, we obtained a sky-projected obliquity of $\lambda = 3.0^{+6.8}_{-4.2}$ deg for TOI-5398 b, consistent with the planet being aligned. With  knowledge of the stellar rotation period, we estimated the true 3D obliquity, finding $\psi = (13.2\pm8.2)$ deg. Based on theoretical considerations, the orientation we measure is unaffected by tidal effects, offering a direct diagnostic for understanding the formation path of this  planetary
system. The orbital characteristics of TOI-5398, with its compact architecture, eccentricity consistent with circular orbits, and hints of orbital alignment, appear more compatible with the disc-driven migration scenario. 

  TOI-5398, with its relative youth (compared with similar compact systems) and exceptional suitability for transmission spectroscopy studies, presents an outstanding opportunity to establish a benchmark for exploring the disc-driven migration model. %through detailed atmospheric characterisation.
}


    %TC:ignore
    \titlerunning{Orbital obliquity of TOI-5398 b}
    \authorrunning{Giacomo Mantovan et al.}
    %TC:endignore

   \maketitle
%
%-------------------------------------------------------------------

\section{Introduction}
\label{sec:introduction}

The obliquity between the orbital axis of a planet and the spin axis of its host star is a key diagnostic for the mechanisms of formation and orbital migration of exoplanets (e.g. \citealt{Naoz2011}). The sky-projected obliquity of a transiting planet can be measured with in-transit radial velocities (RVs) through the Rossiter-McLaughlin (RM) effect (\citealt{Ohta2005}). 

Giant planets in close-in orbits are believed to form in situ close to the final orbit, or in the outer regions and migrate inward \citep{Dawson2018}. Different mechanisms are capable of shrinking the orbits, such as dynamical interactions (high-eccentricity migration scenario) through planet-planet scattering \citep{Marzari2006} or the Kozai mechanism \citep{Wu2003}, and disc--planet interactions \citep{Lin1996}. These alternative mechanisms are expected to imprint different signatures on the obliquity of the planets. Scattering encounters should randomise the alignments of the orbital planes, whereas a migration through disc--planet interactions should keep the planetary orbits roughly coplanar throughout the entire process. 

%Observational results are still inconclusive despite several tens of obliquity measurements. In this context, young transiting planets represent a unique resource for solving this issue. 
Young transiting planets represent a unique resource in their ability to help us interpret yet inconclusive observational results. First, the obliquity may be altered by tidal effects, and close-in planets orbiting stars of a few gigayears old with deep convective envelopes are expected to show aligned configurations regardless of the formation scenario, as is observed in most cases \citep[e.g. ][and references therein]{2022PASP..134h2001A}. Instead, when we observe systems young enough to have avoided tidal alterations of obliquity, we can access the original configuration. Second, a reliable rotation period can be obtained for young, active stars. This enables us to measure the 3D orbital obliquity with respect to the stellar rotation axis while avoiding the bias reported by \cite{2020AJ....159...81M} when the rotation period is included in a Bayesian framework. Third, the amplitude of the RM signature increases with the projected rotational velocity $v \sin{i_\star}$ of the star \citep{winn2010}, more than counterbalancing the adverse effect of rotation and activity on RV errors. Indeed, some planets orbiting active stars have been confirmed by detecting the RM effect rather than via dynamical mass measurements from Keplerian orbit solutions  (e.g. \citealt{2020AJ....160..179M}).

Recent space missions led to the identification of many planets transiting young stars (age $<$ 1 Gyr). The number of confirmed or validated systems is about 70 (cf. NASA Exoplanet Archive). However, the RM effect has only been measured for a limited number of objects, including AU Mic b \citep{2020A&A...643A..25P}, DS Tuc A b \citep{2021A&A...650A..66B}, TOI-942 b \citep{2021ApJ...917L..34W}, V1298 Tau b and c \citep{2022AJ....163..247J,2021AJ....162..213F}, HD 63433 b and c \citep{2020AJ....160..179M,2020AJ....160..193D}, HIP 67522 b \citep{2021ApJ...922L...1H}, TOI-1136 d \citep{2023AJ....165...33D}, and TOI-2076 b \citep{2023ApJ...944L..41F}. These measurements generally indicate aligned configurations, although with moderately large uncertainties, usually ranging between 20 and 30 deg. Consequently, the obliquity of young transiting planets remains significantly unexplored. 

Within the Global Architecture of Planetary Systems (GAPS, \citealt{2013A&A...554A..28C,2020A&A...638A...5C}) long-term program at TNG, many systems with young transiting planets, first identified by space missions, are being characterised through continued RV monitoring (e.g. \citealt{2022A&A...664A.163N,2023A&A...672A.126D}). Among the systems precisely characterised by GAPS, the moderately young solar-analogue star TOI-5398 ($T_{\rm eff} = 6000$K, $R_{\star} = 1.051 \, R_{\odot}$, $M_{\star} = 1.146 \, M_{\odot}$) is of special interest (\citealt{2022MNRAS.516.4432M, 2024A&A...682A.129M}, hereafter \citetalias{2024A&A...682A.129M}). TOI-5398 hosts a hot sub-Neptune planet (TOI-5398\,c, $R_{\rm p} = 3.52 \, R_{\oplus}$, $M_{\rm p} = 11.8 \, M_{\oplus}$, $P \sim 4.77$\,d) orbiting interior to a short-period warm Saturn (TOI-5398\,b, $R_{\rm p}$\,=\,10.30 $R_{\oplus}$, $M_{\rm p} = 58.7 \, M_{\oplus}$, $P \sim 10.59$\,d).

TOI-5398 is the youngest known compact system with a short-period giant, and the giant TOI-5398 b has the highest transmission spectroscopy metric (TSM, \citealt{2018PASP..130k4401K}) value (TSM = 288) among all known warm giant planets (10 < $P$ < 100 d, $M_{\rm p}$ > 0.1 $M_{\rm J}$), making it ideal for atmospheric characterisation with JWST \citepalias{2024A&A...682A.129M}. Given the unusual architecture of the TOI-5398 system and its moderately young age ($650\pm150$\,Myr, $P_{\rm rot} = 7.34$ d), the system serves as a promising target for measuring the original obliquity between the orbital axis of the giant and the spin axis of the star \citepalias{2024A&A...682A.129M}, because it is young enough to have avoided tidal alterations of obliquity. %First, as stated above, we can access the original configuration when observing a system young enough to have avoided tidal alterations of obliquity. 
Moreover, unlike ordinary short-period giants, we can rule out the high-eccentricity migration scenario for compact systems such as TOI-5398 \citep{Mustill2015} and test the other formation and evolution models \citep[e.g. ][]{Lin1996} through detailed atmospheric characterisation. 

Among the compact systems similar to TOI-5398 known to date, namely WASP-47 \citep{2012MNRAS.426..739H}, Kepler-730 \citep{Zhu_2018}, TOI-1130 \citep{Huang_2020}, TOI-2000 \citep{2022arXiv220914396S}, and WASP-132 \citep{2022AJ....164...13H}, only WASP-47 has an obliquity measurement (cf. TEPCat catalogue; \citealt{2011MNRAS.417.2166S,2015ApJ...812L..11S}), and the value obtained is consistent with the system being aligned. Investigating obliquities in multi-planet systems, focusing on uncommon systems such as the ones with an outer close-in giant and an inner small-size planet (listed above), could play a pivotal role in advancing our comprehension of their formation and their relationship to ordinary isolated systems with short-period giant planets \citep{2015ApJ...812L..11S}.

%In this paper, we determine the orbital obliquity of the young planet TOI-5398 b with in-transit RVs (Section \ref{sec:obs}) through the RM effect. In Section \ref{sec:analysis}, we present the procedure used to model the RM effect and estimate the obliquity of the planet, while in Section \ref{sec:discussion}, we discuss our results, provide suggestions for future observations, and highlight dynamical implications for the system's formation and evolution. Concluding remarks are given in Section \ref{sec:conclusion}.

%--------------------------------------------------------------------
\section{Observations and data reduction}
\label{sec:obs}
We observed TOI-5398 with the TNG telescope in the GIARPS configuration \citep{2018SPIE10702E..0ZC} by simultaneously measuring high-resolution spectra in the optical (0.39-0.69 $\mu\textrm{m}$, HARPS-N, $R$ $\sim$ 115 000) and the near-infrared (0.95-2.45 $\mu\textrm{m}$, GIANO-B, $R$ $\sim$ 50 000). In the present work, our analysis is exclusively focused on the HARPS-N spectra. The GIANO-B spectra will be presented in a forthcoming work dedicated to conducting an in-depth atmospheric characterisation of the planet (D'Arpa et al., in prep.).

We collected both in-transit observations  and suitable off-transit observations on March 25, 2023, and obtained a total of 47 spectra with an exposure time of 600 s. This window covers the full 4.3 hour transit (26 spectra) and about 4 hours of off-transit baseline (21 spectra) subdivided into $\sim$ 2 hours before and after the transit event. %We performed the observations within the DDT proposal A46DDT4 (PI: G. Mantovan) allocated time. 
We monitored the latter given the activity level of TOI-5398, for which the typical RV dispersion due to stellar activity on rotation timescale is 29 m s$^{-1}$ \citepalias{2024A&A...682A.129M}.
%Given the activity level of TOI-5398 (typical RV dispersion\textbf{ due to stellar activity} on rotation timescale: 29 m s$^{-1}$, \citetalias{2024A&A...682A.129M}), we monitored the out-of-transit for nearly 4\,hours split in 2\,hours before and after the transit. 
A long out-of-transit observation is required to correct potential RV trends when fitting the RM effect on young stars \citep{2021A&A...650A..66B}. Coincidentally, TOI-5398 c was also transiting during our selected window for TOI-5398 b (see following section). RV measurements of TOI-5398 collected on March 25, 2023 are reported in Table A.1 available at the CDS. %\ref{tab:RVs}. 

We applied the same RV data-reduction procedure outlined in \citetalias{2024A&A...682A.129M} to ensure consistency in our analysis. In particular, we computed the RVs through the cross-correlation function method \citep{2002A&A...388..632P} and using a G2 mask. We obtained RVs with average internal errors $\langle \rm RV_{\rm err} \rangle =$ 5.5 m s$^{-1}$ and average signal-to-noise ratio $\langle \rm S/N \rangle_{5460\AA} =$ 42.

\section{Analysis of the RM effect}
\label{sec:analysis}
Using \texttt{PyORBIT}\footnote{\url{https://github.com/LucaMalavolta/PyORBIT}} \citep{2016A&A...588A.118M, 2018AJ....155..107M}, a Python package designed for modelling planetary transits and RVs while simultaneously accounting for stellar activity effects, we fit the in-transit RV anomaly  in a Bayesian framework following the method introduced by \cite{2013A&A...554A..28C}.

\begin{figure}
   \centering
   \includegraphics[width=\hsize]%{Figures/TOI5398_multidim_RV_i_fixed_RM_vsini_filter_rprs_correct2_b.png}
   {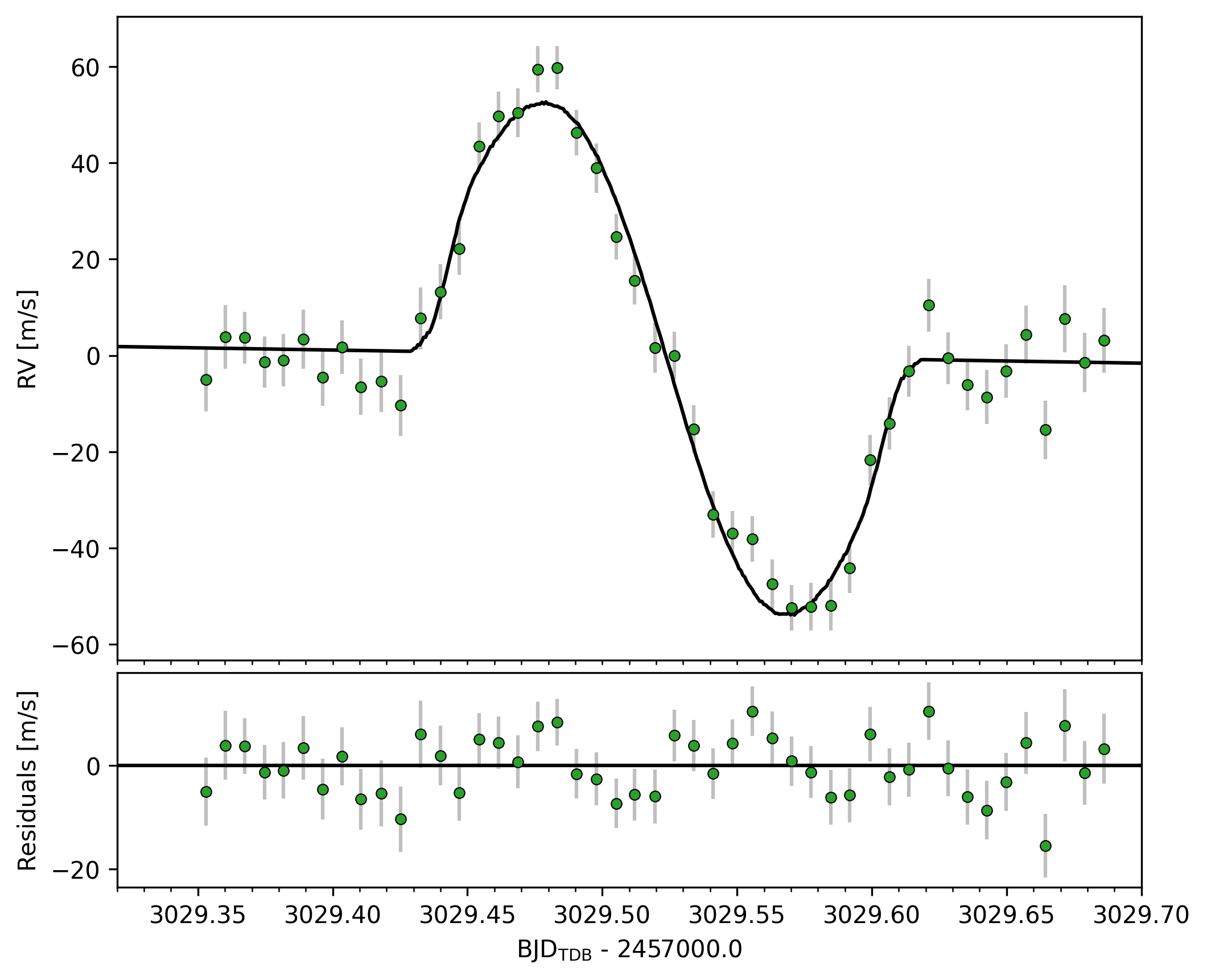}
   \caption{Rossiter-McLaughlin model fit to the RV data collected during the transit of TOI-5398 b. The top panel shows the best-fit RM model (black line) superimposed on the RV observations (green dots). RV data have been corrected for the Keplerian RV curve and systemic RV. The bottom panel displays the residuals resulting from the fit.}
   \label{fig:rm}
\end{figure}

In the model, we relied on stellar and planetary parameters derived in \citetalias{2024A&A...682A.129M} to determine the obliquity of TOI-5398 b. We used priors on orbital and stellar parameters to properly account for their associated errors (see Appendix \ref{app:bayesian_fit} for more details).
We show the data and best-fit RM model in Fig. \ref{fig:rm}, and the best-fitting values of the parameters in Table \ref{table:model-rm}. In particular, from the Bayesian modelling of the RM effect due to planet b, we obtain a sky-projected obliquity of $\lambda\, = 3.0^{+6.8}_{-4.2}$~deg, which is consistent with the planet being aligned.

\begin{table*}
\caption{Priors and outcomes of RM modelling.}             
\label{table:model-rm}      
\centering          
\begin{tabular}{l c c c}     % 7 columns 
\hline\hline     

                      % To combine 7 columns into a single one 
 
\multicolumn{4}{c}{Planet} \rule{0pt}{2.3ex} \rule[-1ex]{0pt}{0pt}\\ 
\hline    
Parameter & Unit & Prior & Value \rule{0pt}{2.3ex} \rule[-1ex]{0pt}{0pt}\\ 
\hline 
   Orbital period ($P_{\rm b}$) & days & $\mathcal{N}$(10.590547, 0.000012) & 10.590547$\pm$0.000012 \rule{0pt}{2.3ex} \rule[-1ex]{0pt}{0pt}\\
   Central time of the first transit ($T_{\rm 0,b}$) & BTJD\tablefootmark{a} & $\mathcal{N}$(2616.49232, 0.00022) & 2616.49232$\pm$0.00022 \rule{0pt}{2.3ex} \rule[-1ex]{0pt}{0pt}\\
   Orbital eccentricity ($e_{\rm b}$) &  & $\mathcal{N}$(0.00, 0.098) & $\leq$ 0.11\tablefootmark{b}%$^{+0.05}_{-0.04}$ 
   \rule{0pt}{2.3ex} \rule[-1ex]{0pt}{0pt}\\
   RV semi-amplitude ($K_{\rm b}$) & m s$^{-1}$ & $\mathcal{N}$(15.7, 1.5) & 15.7$\pm$1.5 \rule{0pt}{2.3ex} \rule[-1ex]{0pt}{0pt}\\
   Planet-to-star radius ratio ($R_{\rm p}/R_{\star}$) &  & $\mathcal{N}$(0.0899, 0.0007) & 0.0898$\pm$0.0007 \rule{0pt}{2.3ex} \rule[-1ex]{0pt}{0pt}\\
   Impact parameter ($b$) &  & $\mathcal{N}$(0.27, 0.07) & 0.270$^{+0.075}_{-0.089}$ \rule{0pt}{2.3ex} \rule[-1ex]{0pt}{0pt}\\
   Omega ($\omega$) & deg & ... & -86$^{+65}_{-67}$ \rule{0pt}{2.3ex} \rule[-1ex]{0pt}{0pt}\\
   Sky-projected obliquity ($\lambda$) & deg & ... & 3.0$^{+6.8}_{-4.2}$ \rule{0pt}{2.3ex} \rule[-1ex]{0pt}{0pt}\\
   True 3D obliquity ($\psi$) & deg & ... & 13.2$\pm$8.2 \rule{0pt}{2.3ex} \rule[-1ex]{0pt}{0pt}\\
\hline

\multicolumn{4}{c}{Stellar parameters} \rule{0pt}{2.3ex} \rule[-1ex]{0pt}{0pt}\\ 
\hline    
Parameter & Unit & Prior & Value \rule{0pt}{2.3ex} \rule[-1ex]{0pt}{0pt}\\ 
\hline 
   Density ($\rho_{\star}$) & $\rho_{\sun}$ & $\mathcal{N}$(0.971, 0.05) & 0.958$^{+0.050}_{-0.049}$ \rule{0pt}{2.3ex} \rule[-1ex]{0pt}{0pt}\\
   Rotation period ($P_{\rm rot}$) & days & $\mathcal{N}$(7.34, 0.05) & 7.34$\pm$0.05 \rule{0pt}{2.3ex} \rule[-1ex]{0pt}{0pt}\\
   Rotational velocity of the star ($v \sin{i_\star}$)\tablefootmark{c} & km s$^{-1}$ & $\mathcal{N}$(7.5, 0.6) & 7.02$^{+0.17}_{-0.21}$ \rule{0pt}{2.3ex} \rule[-1ex]{0pt}{0pt}\\
   Quadratic limb-darkening coefficient ($u_1$) &  & $\mathcal{N}$(0.5760, 0.0149) & 0.574$\pm$0.015 \rule{0pt}{2.3ex} \rule[-1ex]{0pt}{0pt}\\
   Quadratic limb-darkening coefficient ($u_2$) &  & $\mathcal{N}$(0.1377, 0.0318) & 0.135$^{+0.032}_{-0.031}$ \rule{0pt}{2.3ex} \rule[-1ex]{0pt}{0pt}\\
   Stellar inclination ($i_{\star}$) & deg & ... & 76.4$^{+8.3}_{-6.6}$\rule{0pt}{2.3ex} \rule[-1ex]{0pt}{0pt}\\
   Convective blueshift (c$_1$) & & $\mathcal{U}$(-2, 0)& -0.43$^{+0.32}_{-0.59}$\rule{0pt}{2.3ex} \rule[-1ex]{0pt}{0pt}\\
   Stellar equatorial velocity ($v_{\rm eq}$) & km s$^{-1}$ & $\mathcal{U}$(0, 20) & 7.238$\pm$0.100\rule{0pt}{2.3ex} \rule[-1ex]{0pt}{0pt}\\
\hline   
\end{tabular}
\tablefoot{\tablefoottext{a}{BTJD = BJD$_{\rm TDB}$ - 2457000.0.} \tablefoottext{b}{84th percentile.}\tablefoottext{c}{The uncertainty on the modelled $v \sin{i_\star}$ might be slightly underestimated (see e.g. \citealt{2018haex.bookE...2T}).}}
\end{table*}

%\subsection{Simultaneous transit of planet b and c}

\begin{figure}
   \centering
   \includegraphics[width=\hsize]{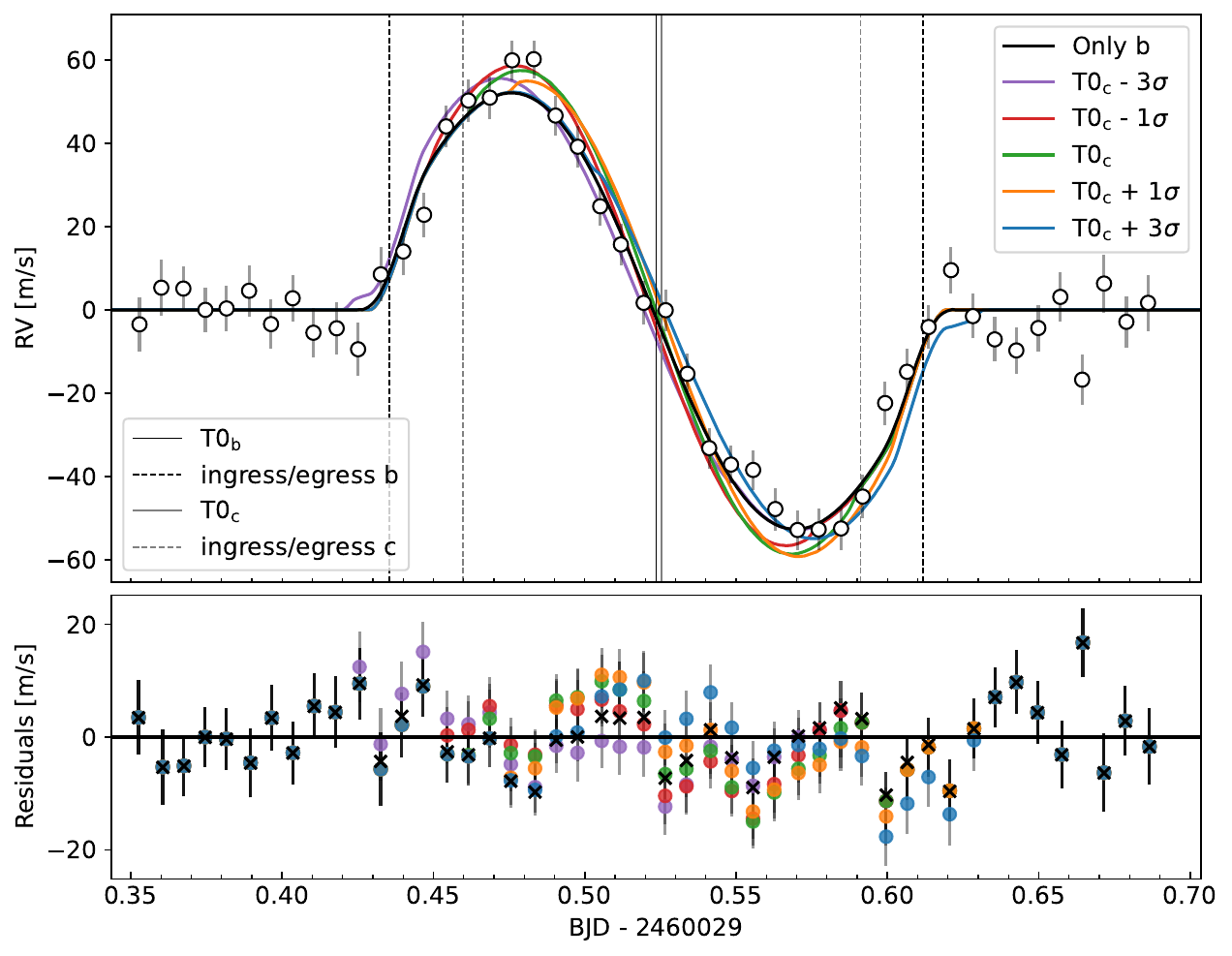}
   \caption{Combined RM effects of planets b and c simulated at different times of inferior conjunction of planet c. For planet c, we assumed null obliquity, while we fixed all the other parameters to the value reported in Table \ref{table:model-rm} and in \citetalias{2024A&A...682A.129M}. For comparison, the RM effect of planet b alone (see Fig. \ref{fig:rm}) is shown as a black line in the upper plot, while black crosses denote the residuals. The vertical lines denote the propagated ephemeris of planets b and c from \citetalias{2024A&A...682A.129M}. %The effect of planet c is on the same order of magnitude as the stellar noise, and it has little effect on the determination of the obliquity of planet b. 
   }
   \label{fig:sim2}
\end{figure}

%Coincidentally, TOI-5398 c was also transiting during our selected window for TOI-5398 b (Fig. \ref{fig:sim2}). 
As briefly mentioned in the previous section, the predicted central transit times of planets b and c (Fig. \ref{fig:sim2}) were very close to each other, implying that the two transits overlapped, even when considering their ephemeris uncertainties (propagated from \citetalias{2024A&A...682A.129M}). We checked how the transit of TOI-5398 c impacts our results. While this degeneracy does not alter our result regarding the alignment of planet b%\footnote{The expected RV anomaly caused by the transit of planet c is an order of magnitude lower than the one caused by the transit of planet b \citepalias{2024A&A...682A.129M}.}
, it slightly increases the associated uncertainty on the measured obliquity (already taken into account).% (Fig. \ref{fig:sim2})
 The current estimate of the uncertainty in the obliquity of planet b can be thought of as an upper limit on the possible contribution to the error budget due to the missing modelling of planet c.
 Indeed, our simulations show that the RM signal associated with planet c has an amplitude similar to the stellar noise observed in our data (expected RV anomaly caused by the transit of planet c\footnote{An order of magnitude lower than the one caused by the transit of planet b, cf. \citetalias{2024A&A...682A.129M}.} $\sim$ 7 m s$^{-1}$ vs. RMS $\sim$ 6 m s$^{-1}$). 

%We performed a preliminary analysis to fit the stellar noise and the RM signals associated with both planets simultaneously, to provide a more robust and precise determination of the obliquity of planet b. The results are compatible, while the obliquity of planet c is unconstrained.

%A preliminary fit including a stellar activity model and the RM signals associated with both planets provided results consistent with those presented in Table \ref{table:model-rm}, while the obliquity of planet c is unconstrained.

%Although we expect the scientific outcome of the analysis to remain the same, we are currently attempting to fit the stellar noise and the RM signal associated with both planets simultaneously to provide a more robust and precise determination of the obliquity of planet b. 

% \begin{figure}[h]
%    \centering
%    \includegraphics[width=\hsize]{Figures/rossiter_toi5398_multi_40kms.pdf}
%    \caption{RV data collected during the transit of planet b, with superimposed the propagated ephemeris of planet b and c from \citetalias{2024A&A...682A.129M}.}
%    \label{fig:sim}
% \end{figure}

\section{Discussion}
\label{sec:discussion}
\subsection{Three-dimensional orbital obliquity}

\begin{figure}
   \centering
   \includegraphics[width=\hsize]{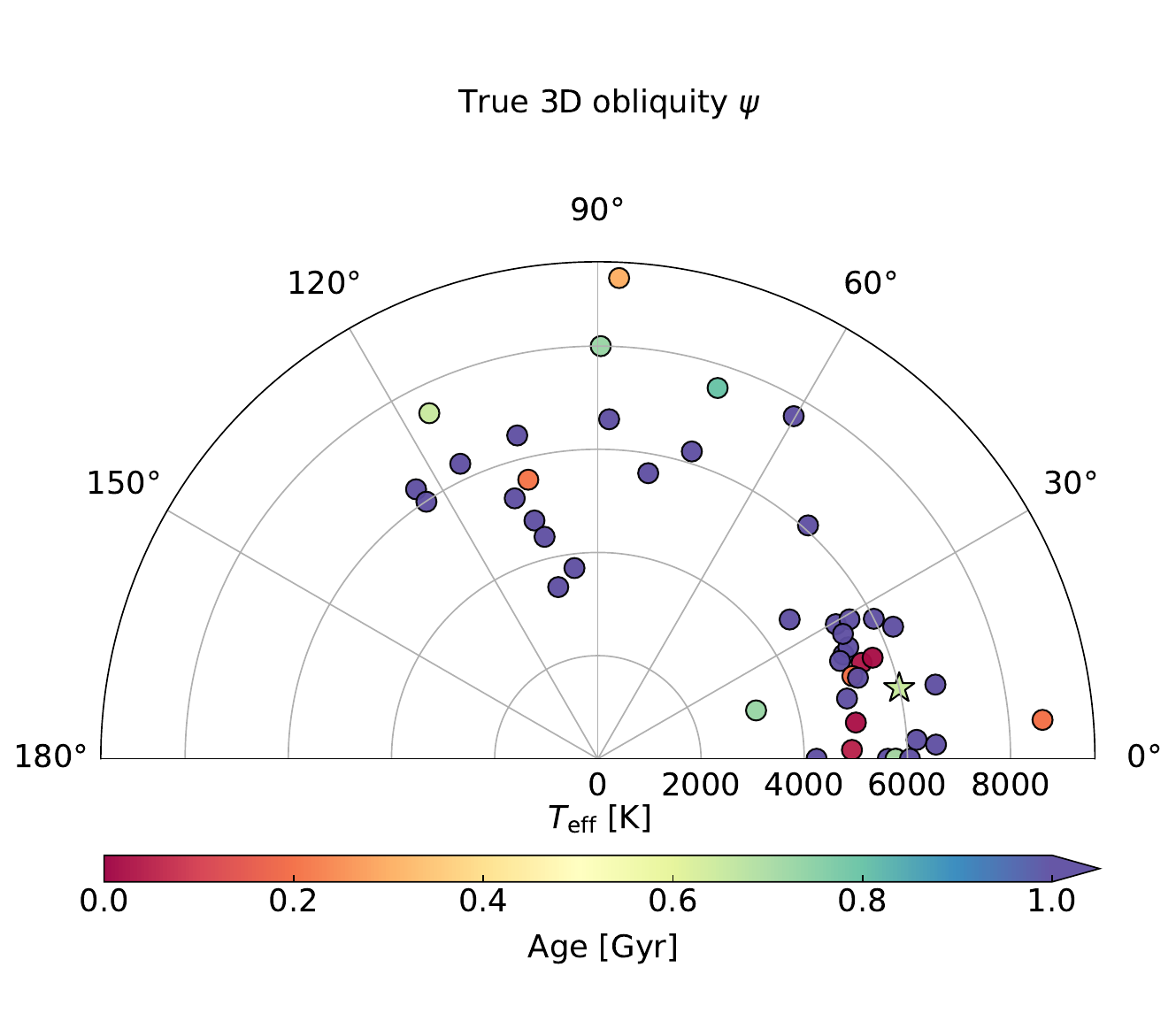}
   \caption{True 3D obliquity of known exoplanets as a function of the effective temperature of the  host star, $T_{\rm eff}$. Each exoplanet except for TOI-5398 (star marker) is represented by a circle. The stellar age is colour coded. The data were taken from TEPCat, NASA Exoplanet Archive, and GAPS (priv. comm.) in January 2024.}
   \label{fig:psi}
\end{figure}

%The best-fitting values of the parameters determined by the analysis of the in-transit RV curve are reported in Table \ref{table:model-rm}. 
%We obtained a projected obliquity $\lambda\, = 2.3^{+4.3}_{-3.7}$ deg, consistent with the planet being aligned. 
Thanks to information on the stellar rotation period ---crucial for probing the stellar inclination $i_\star$ while avoiding the bias on its inference (see \citealt{2020AJ....159...81M})---  and orbital inclination $i_{\rm p}$ provided by \citetalias{2024A&A...682A.129M}, we used Eq. 14 from \cite{2016A&A...588A.127C}:
\begin{equation}
    \psi = \cos{(\sin{i_\star}\cos{\lambda}\sin{i_{\rm p}} + \cos{i_\star}\cos{i_{\rm p}})^{-1}},
\end{equation} and estimated the true 3D obliquity $\psi = (13.2\pm8.2)$ deg. Following the findings from \cite{2022A&A...664A.162M}, which highlight that most exoplanets orbiting cool stars ($T_{\rm eff}$ $<$ 6100~K, known as the Kraft break, \citealt{1967ApJ...150..551K}) tend to have true 3D obliquity $\psi$ $<$ 30 deg, we produced a preliminary polar plot (see Fig. \ref{fig:psi}). This plot shows the 3D obliquity of each known system with a $\psi$ measurement (cf. TEPCat catalogue). The radial coordinate represents the stellar effective temperature $T_{\rm eff}$, while the stellar age is colour coded. The 3D obliquities and $T_{\rm eff}$ come from the TEPCat catalogue, whereas the ages are taken from the GAPS internal database for young stars and from the NASA Exoplanet Archive. It is worth noting that since ages have been estimated using different techniques, they might lack homogeneity. If we focus on the stellar temperature, there appears to be a potential trend around low obliquity values among cool stars (first noted by \citealt{2010ApJ...718L.145W}); though a significant fraction still possess values of between 80 and 125 degrees, as previously noted by \citealt{2021ApJ...916L...1A} and \citealt{2023A&A...674A.120A}. When considering also stellar ages, cool systems ($T_{\rm eff}$ $<$ 6100~K) with obliquities of between 80 and 125 degrees tend to be old (age $>$ 1 Gyr), with the only exception being Kepler-63 b \citep{2013ApJ...775...54S}. All these cool systems with obliquities of between 80 and 125 degrees have scaled distances $a/R_\star >$ 10. Conversely, young cool systems tend to have low obliquity values. Our analysis confirms the trends noted by \cite{2022PASP..134h2001A}, yet the sample is too small to draw conclusions about post-formation misalignment mechanisms.

%In Fig. \ref{fig:3d}, we present the best-fit orbital configuration projection onto the sky plane. The orientation of the sphere has been tilted to show the stellar inclination. We coloured the stellar disc as a function of its surface RV field, which we modelled using \texttt{starry}\footnote{\url{https://github.com/rodluger/starry}} \citep{2019AJ....157...64L} considering the effect of limb darkening. The shaded green area represents the best-fit orbital path, along with a 1$\sigma$ uncertainty range.  

The orbital characteristics of TOI-5398, with its compact architecture, eccentricity values consistent with circular orbits (and stable dynamics; see Appendix \ref{app:dynamics}), and hints of orbital alignment, appear more compatible with the disc-driven migration scenario. This crucial result reveals the potential for exploring the disc-driven migration model through detailed atmospheric characterisation of this system (see Sect. \ref{sec:introduction}).

% \begin{figure}
%    \centering
%    \includegraphics[width=\hsize]{Figures/3d_bis-1.pdf}
%    \caption{Best-fit orbital configuration's projection onto the sky plane. The surface RV field colour-coded the stellar disc, while the white discs represent the planetary orbit.}
%    \label{fig:3d}
% \end{figure}

\subsection{Timescale of the tidal decay of obliquity}
Both the equilibrium tide models of \cite{1977A&A....57..383Z} or \cite{2012ApJ...744..189A} and the dynamical tide model of \cite{2012MNRAS.423..486L} ---this latter having a modified tidal quality factor ($Q^{\prime}_\star$) of $10^5$ for the star\footnote{$Q^{\prime}_\star = (3/2)(Q_{\star}/k_2)$, where $Q_{\star}$ is the tidal quality factor of the star, that is, the ratio of the total maximum energy of its tidal distortion to the energy dissipated by the tides in one cycle, while $k_2$ is the Love number that depends on the internal stratification of the star \citep{2008EAS....29...67Z}.}--- give an obliquity decay timescale exceeding the Hubble time.
%We estimated the timescale of tidal decay of the obliquity according to the approach by \cite{2012MNRAS.423..486L} -- by adopting a modified tidal quality factor of $10^5$ for the star -- and the convective tidal realignment timescale following \cite{1977A&A....57..383Z} and \cite{2012ApJ...744..189A}, finding a decay time exceeding the Hubble time. 
The alignment timescale result is longer than the age of TOI-5398 ($650\pm150$\,Myr), which implies that the obliquity measured should still be unaltered by tidal effects, thus providing a direct diagnostic for the formation path of the planetary system.

We verified the expected tidal realignment timescales for each of the giant planets in the other five compact systems: WASP-47 b, Kepler-730 b, TOI-1130 c, TOI-2000 c, and WASP-132 b. These planets orbit cool stars with convective envelopes, placing them below the Kraft break. Therefore, we can calculate the convective tidal realignment timescale following the same methodology as outlined in \cite{ 2023arXiv230807532W}. %, which warn that the obtained result serves as an order-of-magnitude estimate for the expected realignment timescale rather than a precise value. 
Notably, we find that the decay time greatly exceeds the Hubble time by several orders of magnitude for all checked systems, which confirms that the use of the above approach is well justified. This result shows that RM observations could be used to measure the unaltered obliquities in each of these compact systems and advance our understanding of their formation path.

\subsection{Formation history of the system}
\label{sec:formation}

%To gain deeper insight into the origins of TOI-5398's present architecture we investigated the possible formation and evolution
We investigated the possible formation pathways of the two planets, expanding on the preliminary exploration of \citetalias{2024A&A...682A.129M}. %We first tested if the architecture can be primordial, and thus shaped by disc-driven migration during the planet growth process, by looking for formation tracks resulting in the current planetary masses in the framework of the pebble-accretion scenario. 
Based on the spin--orbit alignment of the system, we assumed the architecture as primordial and shaped by disc-driven migration during the planet growth process. We looked for formation tracks resulting in the current planetary masses in the framework of the pebble-accretion scenario. 
%To this end, we 
We performed Monte Carlo explorations using a modified version of the GroMiT code \citep{polychroni2023}, complementing the growth and migration treatments of solid planets and planetary cores from \cite{johansen2019} and those of gaseous planets from \cite{tanaka2020} with the condensation sequence treatment in protoplanetary discs from \cite{turrini2023} (see Appendix \ref{app:simul_form} for details).

\begin{figure}
   \centering
   \includegraphics[width=\hsize]{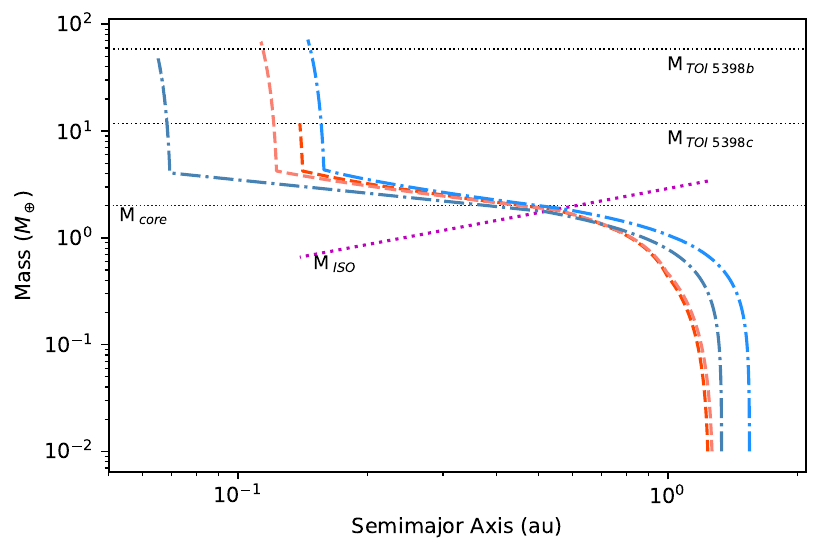}
   \caption{Illustrative example of planetary growth tracks satisfying the selection conditions associated with the black box in Fig. \ref{fig:popsyn} (see main text for details). The growth tracks are projected in the semimajor axis--planetary mass space. The red curves are for planets formed in discs dominated by millimetre-sized pebbles, while the blue curves are for planets formed in discs dominated by centimetre-sized pebbles. All successful tracks are associated with formation regions comprised between 1 and 2\,AU and have planetary cores of about 2\,M$_\oplus$. The core mass is identified by the mass value where the growth tracks intercept the pebble isolation mass curve (purple dotted line).}
   \label{fig:tracks}
\end{figure}

All successful synthetic planets emerge from planetary seeds that formed between 1 and 2\,AU within the first 1\,Myr of the  lifetime of the disc; that is, they grew from first-generation planetesimals born close to the host star (see Fig. \ref{fig:tracks} for illustrative examples). Their formation takes about 2--2.5\,Myr, of which 1.5--2\,Myr are needed for the cores to reach the pebble isolation mass. All these planets are dominated in mass by their gaseous envelopes and possess small cores (about 2\,$M_\oplus$), which means they underwent some level of runaway gas accretion. In the case of planet c, this gas-dominated interior structure appears to be at odds with that expected for planets in its mass range \citep{Fortney2007,LopFor14} and with its modal density value estimated by \citetalias{2024A&A...682A.129M}. However, GroMiT formation tracks provide only lower limits on the interior abundance of heavy elements as they do not yet account for the accretion of planetesimals and high-metallicity disc gas (see e.g. \citealt{turrini2023} for a discussion).% that would increase the amount of heavy elements in the interior of the planet. This is plausibly why it proves comparatively easier to generate synthetic counterparts to planet b than to planet c in our simulations (see Appendix \ref{app:simul_form})

Gas-dominated exoplanets in the mass range of planets b and c show indications of being significantly enriched in heavy elements, enriched by 20--40 times  on average with respect to their host stars \citep{thorngren2016}. Taking into account the slightly supersolar metallicity of TOI-5398 ([Fe/H] = 0.09 $\pm$ 0.06, \citetalias{2024A&A...682A.129M}) and the observed mass--metallicity trend in exoplanets \citep{thorngren2016}, the gaseous envelopes of planets b and c could incorporate about 20 and 6\,$M_\oplus$ of heavy elements, respectively (even more if the envelope of planet c was originally larger; \citetalias{2024A&A...682A.129M}). Due to the limited orbital migration characterising the successful formation tracks, if both planets were formed by pebble accretion, the most likely source for this enrichment is the accretion of high-metallicity gas \citep{booth2019} rather than planetesimals \citep{turrini2021}. 
As suggested by the observations of the Juno mission for Jupiter, such high-metallicity envelopes can mimic the effects of larger cores \citep{wahl2017,stevenson2020}, resulting in an interior structure for planet c more compatible with its density value \citepalias[$\rho_{\rm p} = 1.5$ g cm$^{-3}$,][]{2024A&A...682A.129M}.

Alternatively, if the circumstellar disc of TOI-5398  hosted a significant population of planetesimals, the migrating core of planet b could have accreted the planetesimals along its path and grown by a mixture of pebble and planetesimal accretion without being limited by the pebble isolation mass (see Fig. \ref{fig:tracks}). This would have delayed the onset of its gas accretion \citep{alibert2018,turrini2023}, resulting in a smaller gaseous envelope and the core/envelope ratio characteristic of Neptune-like planets \citep{Fortney2007,LopFor14}.

%To improve our understanding of the photoevaporative evolution of these planets in light of the large uncertainty affecting the density estimation of planet c, we explored a set of end-member scenarios.

\subsection{Photoevaporation history}
\label{sec:photoevap}
The young age of TOI-5398  (650 Myr) stands out among similar compact systems (aged between 3 and 10 Gyr), making it a prime target for photoevaporation studies. We evaluated the mass-loss rate of the planetary atmospheres with the new ATES photoionisation hydrodynamics code \citep{Caldiroli+2021,Caldiroli+2022}, coupled with the planetary core--envelope models by \cite{Fortney2007} and \cite{LopFor14}, the MESA Stellar Tracks (MIST; \citealt{choi+2016}), and the extreme-ultraviolet (XUV) luminosity time evolution \citep{SF22}. To perform simulations of the past evolution of the planetary atmospheres, we determined the planetary core mass and radius of the planets in TOI-5398 at their present age (details in Appendices \ref{app:structure} and \ref{app:photoevap}). The best agreement with the measured values was obtained by assuming high-metallicity atmospheric envelopes. %These parameters are supposed to remain constant over time. 

Specifically, assuming cores composed of 50\% rocks and 50\% ices, we find core sizes larger than predicted by the formation models if simple pebble accretion is assumed. Higher initial core/envelope ratios could be obtained if we were to postulate dense high-metallicity envelopes or pebble+planetesimal accretion scenarios (as proposed in Sect.\ \ref{sec:formation}).
Alternatively, we were able to find configurations with smaller rock--iron cores (i.e. Earth-like core composition with 67\% rocks and 33\% iron) and very large envelopes at the present age of the system, in line with the formation scenario, but only for planet b. 
 
We investigated the hydrodynamic atmospheric evolution of the two planets, from $t_0 = 3$\,Myr (i.e. the end time of planetary formation models) to 650\,Myr. We performed several simulations for both planets, assuming cores of different sizes.
% either the relatively large cores derived in Appendix\,\ref{app:structure} or the smaller cores suggested by the planetary formation models (Sect.\ \ref{sec:formation}). 
Our results, which are  shown in Figs. \ref{fig:evaporation_mass_radius}, \ref{fig:evaporation_atmo}, and in Table \ref{table:evpar},  confirm that planet b experienced a relatively minor atmospheric escape during its lifetime, and it remains dominated by a massive and extended envelope, in agreement with the GroMiT results (Sect. \ref{sec:formation}). Assuming a small or a large core does not affect the evolution appreciably. The reconstructed planetary radius at $t_0$ was a factor 2.2--2.4 times the value at the current age.
%, and it will decrease by just $\sim 15$\% until an age of 5\,Gyr.

In the case of planet c, the evolution was more important. In particular, assuming a large core, the atmospheric mass fraction decreased from $\sim 38$\% at $t_0$ to 1.5\% at the present age; hence, the initial planetary mass was $\sim 60$\% larger than the current value. 
The mass-loss rate was a factor $\sim 6$--7 higher than for planet b at $t_0$, but now it is slightly lower. This difference can be explained by considering that the XUV irradiation remained a factor $\sim 3$ higher during the evolution of planet c, while its mean density increased from 0.01\,g cm$^{-3}$ at $t_0$ to 1.5\,g cm$^{-3}$ today due to a decrease in the planet radius by a factor 5.8. We cannot find a reasonable evolutionary path for planet c if a small core is assumed because the high efficiency of the photoevaporation in the past implies a relatively thin atmospheric envelope at the present age, and therefore the core size cannot be much smaller than the planet. 
% Nevertheless, we confirm that planet c will lose entirely its residual atmosphere in less than 250\,My. 

\section{Conclusions}
\label{sec:conclusion}

%\com{WE ARE AT 3006 WORDS IN THIS MOMENT (MENU -> WORD COUNT; appendix/acknowledgements not included)}

We conclude by outlining the key findings from this study:
\begin{enumerate}
    \item The system's orbital plane is well aligned with the equatorial plane of the star, indicating its architecture is %primordial and 
    shaped by disc-driven migration.
    \item The orientation we measure is unaffected by tidal effects, meaning that a direct diagnostic of the
formation path is possible  for this planetary system.    \item Planet b experienced little mass loss. Hence, the current size and structure of this gas giant are likely primordial. Its current radius suggests an envelope with enhanced opacity. 
    %We predict the existence of an important atmosphere; 
    %, which can explain the deep He I absorption signal probed with transmission spectroscopy (D'Arpa et al., in preparation).
    \item Planet c was more significantly shaped by photoevaporation. Its modal density \citepalias{2024A&A...682A.129M} is suggestive of an atmosphere and appears incompatible with the small core set by pebble accretion alone, as the efficient photoevaporation would have shrunk the planetary radius beyond what is observed. %The current uncertainty on the planet mass, however, does not allow to draw stronger conclusions;
    \item The presence of a large core inside planet c argues in favour of formation in a planetesimal-rich disc, while a small core surrounded by high-metallicity gas favours a pebble-rich disc. As both planets were born in the same circumstellar disc, these two formation scenarios also have different implications for the composition of planet b.
    \item Future observations with JWST can help discriminate between the two scenarios; for example by constraining the carbon--sulphur ratio \citep{turrini2021,crossfield2023}: a large C/S ratio would point to a pebble-rich disc, while a lower one to a planetesimal-rich disc.
    %was affected by substantial photoevaporation. Hence, it now retains an atmospheric envelope much thinner than that of planet b. The higher density also implies a slightly lower mass loss rate at present age.
    %\item The adopted planetary structure models suggest larger cores than formally indicated by planetary formation models. For planet b, such models also predict a planetary radius smaller than measured (within $2\sigma$) and/or a higher planetary mass.
    \item TOI-5398's youth (650 $\pm$ 150 Myr) stands out among similar compact systems (aged between 3 and 10 Gyr), making it an exceptional target for photoevaporation studies. The possible survival of a primordial atmosphere of the inner companion is crucial for yielding comparative planetology and exploring the disc-driven migration model through detailed atmospheric characterisation.
    
    %-> not "frozen" like the other systems but young and "evolving" (photoevaporation still in progress) -> may offer new answers to the evolution of such systems (and short-period giant planets in general??) -> inner companion's atmosphere might still be primordial -> essential to check disc-driven migration!!!

 \end{enumerate}

%\color{purple}
%NOTA per le conclusioni da parte di DT \& DP: come discusso nel testo che abbiamo preparato, la migrazione molto limitata prevista per questi pianeti favorisce il gas del disco come fonte della metallicità del loro inviluppo rispetto all'accrescimento di planetesimi. Una possibile implicazione osservativa di questo scenario per JWST è di una atmosfera molto arricchita in elementi volatili (C e O) e limitatamente arricchita in elementi refrattari come lo zolfo.
%\color{black}

%TC:ignore   
\begin{acknowledgements}
We greatly acknowledge the fundamental support provided by the TNG director Adriano Ghedina in approving the allocated time for DDT proposal A46DDT4 (PI: G. Mantovan). For the same reason, we also acknowledge the support provided by Vania Lorenzi, Aldo Fiorenzano, and Walter Boschin. GMa acknowledges support from CHEOPS ASI-INAF agreement n. 2019-29-HH.0. This work benefited from the 2023 Exoplanet Summer Program in the Other Worlds Laboratory (OWL) at the University of California, Santa Cruz, a program funded by the Heising-Simons Foundation and NASA. DP acknowledges the support from the Istituto Nazionale di Oceanografia e Geofisica Sperimentale (OGS) and CINECA through the program ``HPC-TRES (High Performance Computing Training and Research for Earth Sciences)'' award number 2022-05 as well as the support of the  ASI-INAF agreement n 2021-5-HH.1-2022. AMa acknowledges partial support from the PRIN-INAF 2019 (project HOT-ATMOS). RSp acknowledges the support of the ARIEL ASI/INAF agreement n. 2021-5-HH.0. RSp, GMi acknowledge the support of the ASI-INAF agreement n 2021.5-HH.1-2022. This work has been supported by the PRIN-INAF 2019 ``Planetary systems at young ages (PLATEA)''. The authors acknowledge financial contribution from PRIN MUR 2022 (project No.\,2022CERJ49), PRIN-INAF 2019, and the ASI-INAF agreement n.2018-16-HH.0 (THE StellaR PAth project). GPi, LMa acknowledge partial support by DOR2023 DFA, UNIPD. L.M. acknowledges financial contribution from PRIN MUR 2022 project 2022J4H55R. TZi acknowledges NVIDIA Academic Hardware Grant Program for the use of the Titan V GPU card and the support by the CHEOPS ASI-INAF agreement n. 2019-29-HH.0 and the Italian MUR Departments of Excellence grant 2023-2027 ``Quantum Frontiers''.
\end{acknowledgements}
%TC:endignore


\bibliographystyle{aa} % style aa.bst
\bibliography{references} % your references 

\begin{appendix}
%TC:ignore
\section{Radial velocities}
The RV spectroscopic time series will be available in electronic format as supplementary material for the Letter at the CDS.
%are included in Table \ref{tab:RVs}.
%\input{table/longtable}

\section{Bayesian fit}
\label{app:bayesian_fit}
To properly take into account the errors associated with the orbital parameters, we used Gaussian priors on the orbital period ($P_{\rm b}$), the central time of the first transit ($T_{0,{\rm b}}$), the planet-to-star radius ratio ($R_{\rm b}/R_{\star}$), and the impact parameter ($b$). We imposed a Gaussian prior on the eccentricity following \cite{2019AJ....157...61V} and included the RV semi-amplitude $K_{\rm b}$ in the modelling. We also imposed Gaussian priors on the host star density ($\rho_{\star}$), rotational period ($P_{\rm rot}$), and projected rotational velocity ($v \sin{i_\star}$). In particular, we treated the stellar equatorial velocity ($v_{\rm eq}$, obtained from $P_{\rm rot}$) and inclination ($i_\star$) as independent variables. Therefore, we are not subject to the bias described in \cite{2020AJ....159...81M}. We treated the stellar limb-darkening (LD) contribution by estimating $\rm u_1$ and $\rm u_2$ using \texttt{PyLDTk}\footnote{\url{https://github.com/hpparvi/ldtk}} \citep{2013A&A...553A...6H, 2015MNRAS.453.3821P} and assuming a boxcar filter as the passband in the HARPS-N spectral range. We also modelled and fit the effect of the stellar convective blueshift (CB) on the in-transit RV curve \citep[e.g. ][]{2016A&A...588A.127C} using a linear law as a function of the limb angle. Short-term stellar activity was included in the model through a jitter term to be added in quadrature to the errors of the RV.

We performed a global optimisation of the parameters by running \texttt{PyDE} \citep{Storn1997} and performing a Bayesian analysis of the RM signal in the RV time series using \textsc{emcee} \cite{2013PASP..125..306F}. This choice was motivated by the fact that MCMC samplers allow the specification of a prior on derived parameters, as in the case of $v \sin{i_\star}$, which is not possible in the case of nested sampling algorithms.

We performed a preliminary analysis including a  Gaussian Process with a Matern-3/2 kernel \citep[e.g.][]{2020A&A...643A..25P} to model the stellar activity, and the simultaneous modelling of the RM signals associated with both planets. The results are perfectly consistent with those presented in Table \ref{table:model-rm} with comparable error bars, while the obliquity of planet c remains unconstrained. For this reason, we preferred the simpler model presented in Sect. \ref{sec:analysis}. A more complex modelling will be presented after securing more spectroscopic and photometric in-transit data.

%\newpage
\section{Simulating planet-formation tracks}
\label{app:simul_form}
\begin{figure}
   \centering
   \includegraphics[width=\hsize]{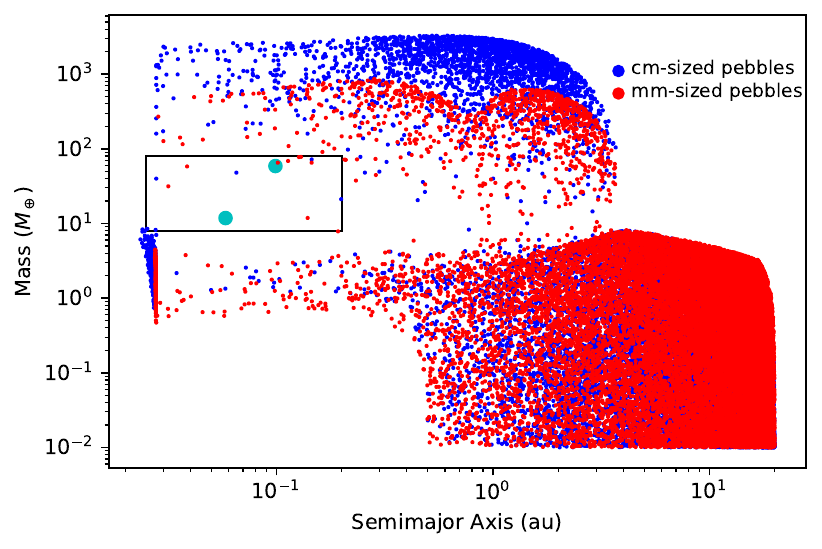}
   \caption{Synthetic populations of planets resulting from the two Monte Carlo runs of 20000 extractions each with GroMiT. The plot shows the final masses and orbital positions of the simulated growth tracks (see Fig. \ref{fig:tracks} for illustrative examples of the growth tracks). The red symbols are associated with planets formed in discs dominated by mm-sized pebbles, and the blue symbols with planets formed in discs dominated by cm-sized pebbles. TOI-5398 b and c are represented by the two larger cyan-filled circles. The black box highlights the region of the parameter space populated by the successful solutions (see main text for details).}
   \label{fig:popsyn}
\end{figure}

We considered a template protoplanetary disc similar to the nominal one from \cite{turrini2023} (see Table \ref{tab:popsythesis} for a summary of its main physical parameters) and sampled 4$\times$10$^4$ planet formation tracks. 
%This modified version of GroMiT the flux of pebbles accreted by the growing planet across the different regions of the protoplanetary disc using the simplified condensation sequence of the ices from \cite{turrini2023}. 
%We sampled 2$\times$10$^4$ formation tracks assuming cm-sized pebbles and 2$\times$10$^4$ tracks assuming
Half of our samples assume centimetre(cm)-sized pebbles and the other half millimetre(mm)-sized pebbles, to explore the impact of faster and slower growth rates of the planetary cores \citep{johansen2019}. The resulting populations of synthetic planets are shown in Fig. \ref{fig:popsyn}. Each formation track is characterised by a different combination of the initial time and orbital distance of the planetary seed, both values being obtained using uniform extractions (see Table \ref{tab:popsythesis} for the relevant ranges and see Fig. \ref{fig:tracks} for illustrative examples of the growth tracks).

Because of the uncertainties of the real characteristics and lifetime of TOI-5398's now-dispersed disc, we considered as successful all planets between 8 and 80 $M_\oplus$, that is roughly between 20\% lower than the mass of planet c and 20\% higher than the mass of planet b at their current age, and whose final orbits fall within 0.2\,AU. This region in the space of the solutions is identified by the black box in Fig. \ref{fig:popsyn}. Consistently with \cite{johansen2019}, we find that forming planets in this mass range is more challenging than forming less massive super-Earths ($<5$\,$M_\oplus$) and more massive gas giants ($>100$\,$M_\oplus$), and that it is comparatively easier to form synthetic counterparts of planet b than of planet c (see Fig. \ref{fig:popsyn}).% When we account for the preliminary photoevaporation investigation performed by \citetalias{2024A&A...682A.129M}, the latter result argues in favour of planet c having been more massive in the past and having undergone stronger mass loss than planet b.

\begin{table}[h!]
    \caption{Input parameters used to run the modified GroMiT code to produce simulated planets for the TOI 5398 system's planets b and c. }
    \label{tab:popsythesis}
    \centering
    \begin{tabular}{l c c}
        \hline \hline
%        \multicolumn{1}{c}{Parameter} & Simulation 1 & Simulation 2 \\
%        \hline
    \multicolumn{2}{c}{Simulation Parameters} \rule{0pt}{2.3ex} \rule[-1ex]{0pt}{0pt}\\
    \hline
      \multicolumn{1}{l}{N$^\circ$ of simulations} & \multicolumn{1}{c}{4$\times10^4$} \rule{0pt}{2.3ex} \rule[-1ex]{0pt}{0pt}\\
      \multicolumn{1}{l}{Seed formation time} & \multicolumn{1}{c}{0.1--1.0$\, \times \, 10^6$\,yr} \\
      \multicolumn{1}{l}{Disc lifetime} & \multicolumn{1}{c}{$3.0 \times 10^6$\,yr} \\
    \hline
    \multicolumn{2}{c}{Star, planet, and disc properties} \rule{0pt}{2.3ex} \rule[-1ex]{0pt}{0pt}\\
    \hline
      \multicolumn{1}{l}{Stellar mass} & \multicolumn{1}{c}{1.15$\,$M${_\odot}$} \rule{0pt}{2.3ex} \rule[-1ex]{0pt}{0pt}\\
      \multicolumn{1}{l}{Seed mass} & \multicolumn{1}{c}{0.01$\,$M${_\oplus}$} \\
      \multicolumn{1}{l}{Initial envelope mass} & \multicolumn{1}{c}{0.0$\,$M${_\oplus}$} \\
      \multicolumn{1}{l}{Initial semimajor axis} & \multicolumn{1}{c}{0.5--20.0$\,$au} \\
      \multicolumn{1}{l}{Disc characteristic radius} & \multicolumn{1}{c}{50.0$\,$au} \\
      \multicolumn{1}{l}{Temperature at 1$\,$au} & \multicolumn{1}{c}{200$\,$K} \\
      \multicolumn{1}{l}{Surface density at 1$\,$au} & \multicolumn{1}{c}{420\,kg m$^{-2}$} \\
      \multicolumn{1}{l}{Disc accretion coefficient, $\alpha$} & \multicolumn{1}{c}{0.01}\\
      \multicolumn{1}{l}{Turbulent viscosity, $\alpha$$_\nu$}& \multicolumn{1}{c}{0.0001}\\
      \multicolumn{1}{l}{Pebble size\tablefootnote{Half of the simulations where run using pebble sizes of 1$\,$cm and the other half 1$\,$mm.} }& \multicolumn{1}{c}{1$\,$cm \& 1$\,$mm}\\
    \hline
  \end{tabular}
%  \tablefoot{\tablefoottext{a}{}}
\end{table}

\section{Planetary structure}
\label{app:structure}

\begin{figure*}[]
\centering
\subfigure[]{\includegraphics[width=0.45\textwidth, trim = 2cm 0cm 2cm 0.2cm]{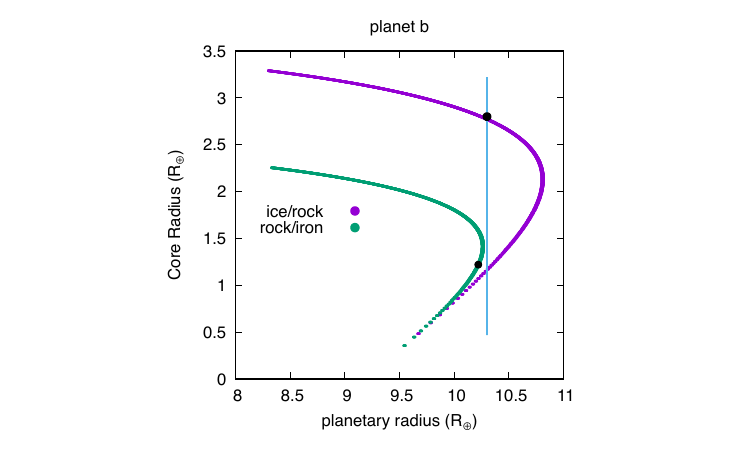}
% {rcore_radius1.pdf}
}
\subfigure[]{\includegraphics[width=0.45\textwidth, trim = 2cm 0cm 2cm 0.2cm]{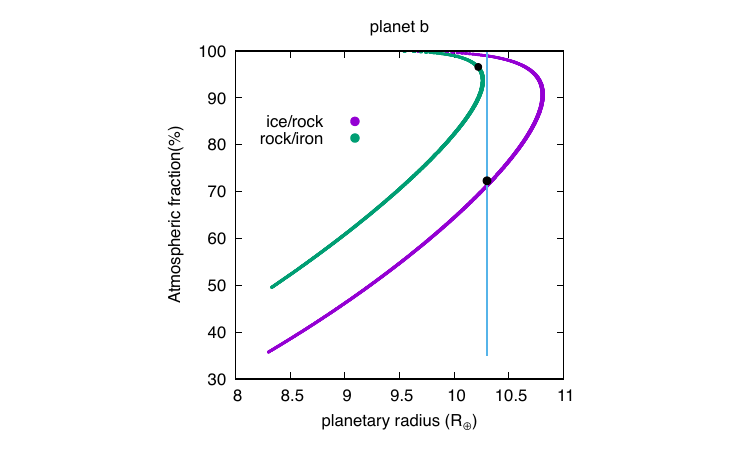}
% {massfractio_radius1.pdf}
}
\caption{Solutions of core--envelope models for planet TOI-5398\,b. (Left) Core radius vs.\ theoretical planetary radius for total mass fixed at the measured value and a rock-ice core (locus of purple points) or a rock-iron core (green points). 
The vertical line indicates the measured planetary radius.
The two black points mark two different configurations selected for exploring the photoevaporation history of the planet. (Right) Similar to the previous plot but for the atmospheric mass fraction.} 
\label{fig:coreb}
\end{figure*}

To determine the atmospheric mass fraction at present age, we adopted the core-envelope structure by \cite{Fortney2007} and \cite{LopFor14}. This core-envelope model has four unknowns: the mass and radius of the core, $M_{\rm core}$ and $R_{\rm core}$, the radius of the envelope, $R_{\rm env}$, and the atmospheric mass fraction, $f_{\rm atm}$. These quantities are linked by four relations and two observational constraints: the measured planet radius and total mass, $R_{\rm p}$ and $M_{\rm p}$; the \cite{LopFor14} relation which links $R_{\rm env}$, $f_{\rm atm}$, and $M_{\rm p}$; and a relation between $R_{\rm core}$, $R_{\rm p}$, and $M_{\rm p}$ at the given age and star--planet distance. For the latter, we adopted the internal structure models by \cite{Fortney2007}, where cores can be composed of different ice-rock or rock-iron mixtures.

We note that the relation by \cite{LopFor14} was developed for H--He-dominated atmospheres, and it also takes into account the cooling and contraction of the envelope as a consequence of its thermal evolution \citep{LopForMil12}. For atmospheres with solar metallicity, the radius of the envelope decreases with age as $t^{-0.11}$, while for enhanced opacities the variation is $\propto t^{-0.18}$.

Assuming cores composed of 50\% rocks and 50\% ices, and a high-metallicity envelope, we found for planet b a first closed solution with $M_{\rm core,b} = 16.9$\,\mearth\ and $R_{\rm core,b} = 2.8$\,\rearth, corresponding to the measured planetary mass and optical radius (Fig.\ref{fig:coreb}). The alternative assumption of solar metallicity yields a maximum value of the planetary radius that is about 7\% lower than the available measurement. The core and envelope densities result around $4.2$\,g cm$^{-3}$ and $0.55$\,g cm$^{-3}$, respectively. The mean planetary density is around $0.3$\,g cm$^{-3}$.
% Note that in Mantovan et al. (2023) we had explored solutions with core radii in the range 3--7 \rearth for planet b, having corresponding core masses 10--25 \mearth, while for planet c we had limited our analysis to core radii in the range 1--2.5 \rearth and core masses $\sim 10$ \mearth.

For planet c, we obtained $M_{\rm core,c} = 11.7$\,\mearth\ and $R_{\rm core,c} = 2.6$\,\rearth, a solution self-consistent with the assumed core-envelope structure and the measured mass and optical radius. These values imply a core density of $\sim 3.6$\,g cm$^{-3}$, an envelope density $\sim 1.3$\,g cm$^{-3}$, and a mean planetary density $\sim 1.5$\,g cm$^{-3}$. These core sizes and masses are larger than predicted by the planetary formation models described in Sect.\ \ref{sec:formation}, while the densities are lower.

Alternatively, we searched for a second possible configuration assuming a denser Earth-like core composition with 67\% rocks and 33\% iron. In this case, we found a solution only for planet b with a smaller and less massive core (Fig.\ref{fig:coreb}), having $M_{\rm core,b} = 2.0$\,\mearth\ and $R_{\rm core,b} = 1.2$\,\rearth, more in line with the prediction of the planetary formation models.

\begin{figure*}
\centering 
\subfigure[]{\includegraphics[width=0.45\textwidth, trim = 1cm 13cm 2cm 3cm]{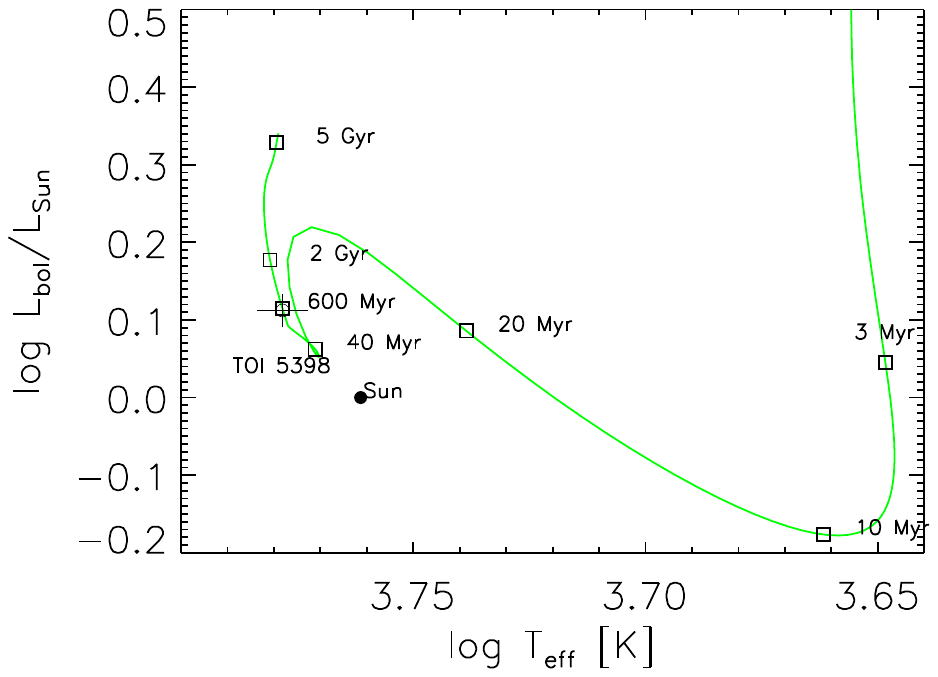}
}
\subfigure[]{
\includegraphics[width=0.45\textwidth, trim = 1cm 13cm 2cm 3cm]{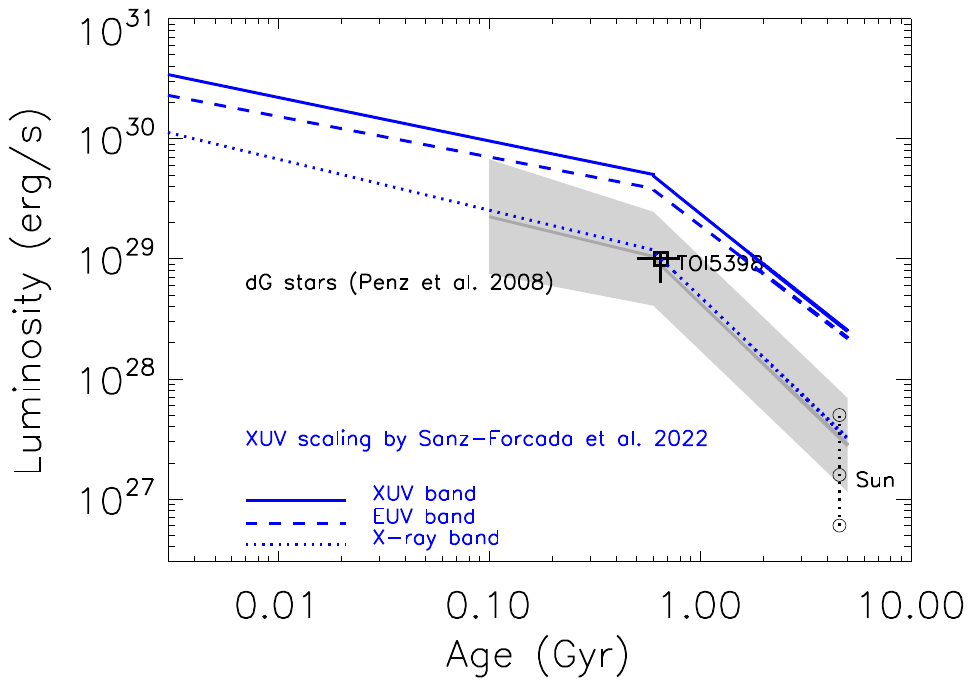}
}
\caption{
(Left) Evolutionary track of TOI\,5398 up to an age of 5 Gyr in the effective temperature--bolometric luminosity plane. A black cross marks the current location of the star on the track.
(Right) Time evolution of X-ray (5--100\,\AA), EUV (100--920\,\AA), and total XUV luminosity of TOI\,5398, according to \cite{Penz08a} and the X-ray/EUV scaling by 
\citet{SF22}.
The grey area is the observed $1\sigma$ spread around the median (dark grey line) of the X-ray luminosity distributions for dG stars in the Hyades and Pleiades open clusters. Uncertainties on the age and X-ray luminosity of TOI\,5398 are also indicated.
}
\label{fig:evoltrack}
\end{figure*}

% \begin{figure}
% \centering
%       \subfigure[]{
% \includegraphics[width=0.45\textwidth, trim = 1cm 13cm 2cm 3cm]{LXUV_time_pm08jo21_TOI5398.pdf}
%}
%    \subfigure[]{
% \includegraphics[width=0.45\textwidth, trim = 1cm 13cm 2cm 3cm]{LXUV_time_pm08s22_TOI5398.pdf}
%}    
% \caption{Time evolution of X-ray (5--100\,\AA), EUV (100--920\,\AA), and total XUV luminosity of TOI\,5398, according to \cite{Penz08a} and the X-ray/EUV scaling by 
% \citet{Johnstone+2021} (left panel) or 
% \citet{SF22}. % (right panel). 
% The grey area is the observed $1\sigma$ spread around the median (dark grey line) of the X-ray luminosity distributions for dG stars in the Hyades and Pleiades open clusters. Uncertainties on the age and X-ray luminosity of TOI\,5398 are also indicated.}  
% \label{fig:xuvevol}
% \end{figure}

\section{Photoevaporation modelling}
\label{app:photoevap}
We adopted the ATES photoionisation hydrodynamics code by \cite{Caldiroli+2021,Caldiroli+2022}.
To speed up the computation, we resorted to their analytical approximation, which provides the planetary mass-loss rate as a function of planetary mass and radius, star--planet distance, and stellar high-energy flux. This is achieved by evaluating the effective photoevaporation efficiency, $\eta_{\rm eff}$, which enters into the classical energy-limited formulation \citep{Erkaev+2007}:
\begin{equation}
    \Dot{M} = \eta_{\rm eff} \frac{3 F_{\rm XUV}}{4 G K \rho_{\rm p}}~,
    \label{eq:mdot}
\end{equation}
where $F_{\rm XUV}$ is the X-EUV flux at the (average) orbital distance,
$\rho_{\rm p}$ is the mean planetary mass density, and the factor K accounts
for the host star tidal forces \citep{Erkaev+2007}. In all cases, we consider the stellar evolutionary track and the long-term change of the XUV irradiation.

The track followed by TOI-5398 in the theoretical temperature--luminosity diagram is shown in Fig. \ref{fig:evoltrack}, and was computed using the web-based interpolator\footnote{\url{https://waps.cfa.harvard.edu/MIST/interp_tracks.html}} of the MESA Isochrones and Stellar Tracks (MIST, \citealt{choi+2016}).

\begin{figure*}[]
\centering
\subfigure[]{\includegraphics[width=0.45\textwidth, trim = 2cm 0cm 2cm 0.2cm]{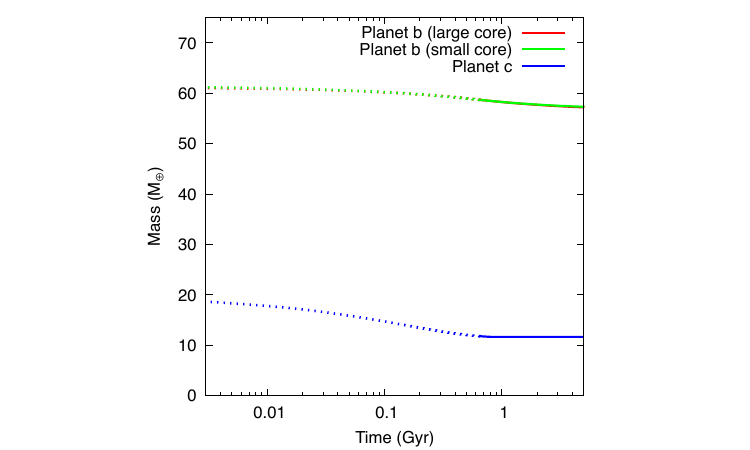}}
\subfigure[]{\includegraphics[width=0.45\textwidth, trim = 2cm 0cm 2cm 0.2cm]{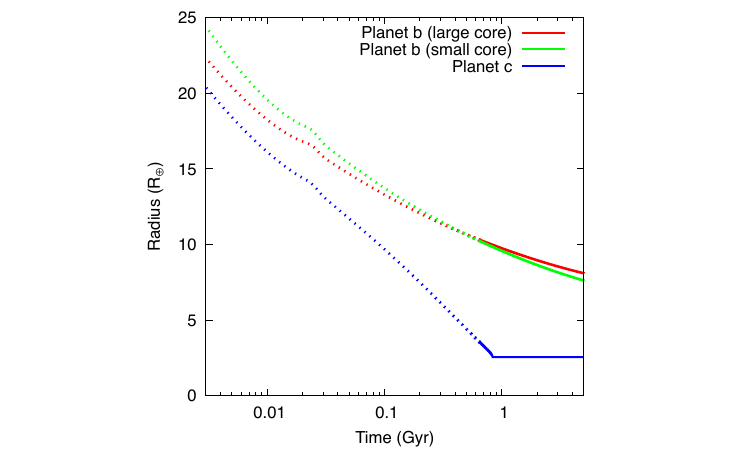}}
\caption{Evolutionary history of TOI-5398 planets. Panels (a) and (b) show the planetary mass and radius vs. time. Dotted segments indicate the evolution backwards from the current age and solid lines evolution forward in time. These simulations were constructed with the initial conditions given by the planetary mass and radius measured at the present age of 650\,Myr.}     
\label{fig:evaporation_mass_radius}
\end{figure*}
        
\begin{figure*}[]
\centering
\subfigure[]{\includegraphics[width=0.45\textwidth, trim = 2cm 0cm 2cm 0.2cm]{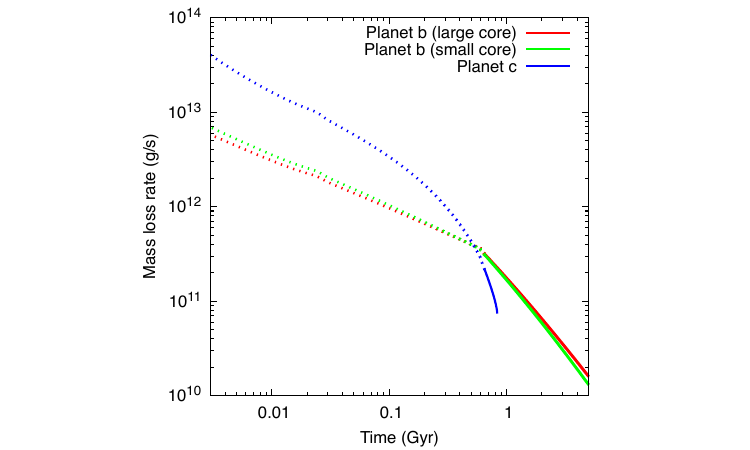}}
\subfigure[]{\includegraphics[width=0.45\textwidth, trim = 2cm 0cm 2cm 0.2cm]{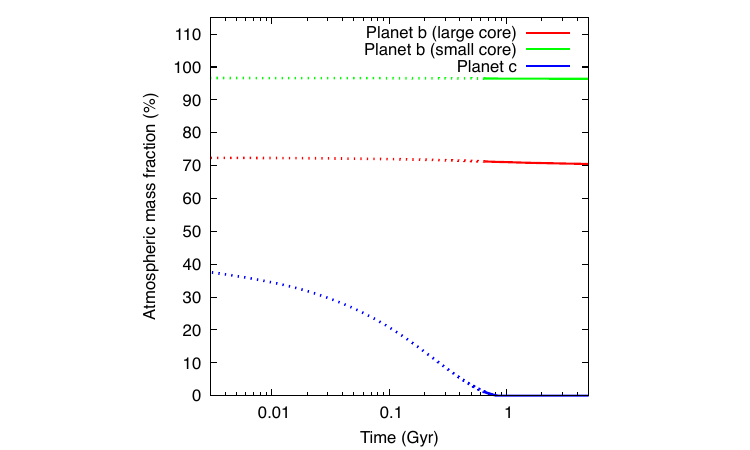}}
\caption{Similar to Fig.\ \ref{fig:evaporation_mass_radius} but for the time evolution of the mass-loss rate (a) and the atmospheric mass fraction (b).
} 
\label{fig:evaporation_atmo}
\end{figure*}

\begin{table*}
    \caption{Photoevaporation modelling outcome.}
    \label{table:evpar}     \centering              \begin{tabular}{c c|c c c c|c c c c}         
        \hline \hline              Core Radius & Core Mass & Mass & Radius & $f_{\rm atm}$ & Mass loss rate & Mass & Radius & $f_{\rm atm}$ & Mass loss rate \rule{0pt}{2.3ex} \rule[-1ex]{0pt}{0pt}\\     
    (\rearth) & (\mearth) & ($M_\oplus$) & ($R_\oplus$) & (\%) & (g/s)  & ($M_\oplus$) & ($R_\oplus$) & (\%) & (g/s) \rule{0pt}{2.3ex} \rule[-1ex]{0pt}{0pt} \\    
    \hline
    
    \multicolumn{2}{c}{Planet b} & \multicolumn{4}{c}{current age} & \multicolumn{4}{c}{at 3\,Myr} \rule{0pt}{2.3ex} \rule[-1ex]{0pt}{0pt}\\
    \hline
%   2.2 & 6.5 & 58.7 & 9.7 & 88.9 & $2.7 \times 10^{11}$ & 60.4 & 15.4 & 89.2 & $2.8 \times 10^{12}$ \\
    2.8 & 16.9 & 58.7 & 10.3 & 71.3 & $3.2 \times 10^{11}$ & 61.1& 22.3 & 72.4 & $5.8 \times 10^{12}$ \rule{0pt}{2.3ex} \rule[-1ex]{0pt}{0pt}\\
    1.2 & 2.0 & 58.7 & 10.2 & 96.6 & $3.1 \times 10^{11}$ & 61.2& 24.4 &    96.7 & $6.9 \times 10^{12}$ \\

%    1.2 & 1.8 & 60.0 & 9.0 & 97.0 & $8 \times 10^{11}$ & 61.2 & 15.2 & 97.0 & $4 \times 10^{11}$ \\

    \hline                       
    \multicolumn{2}{c}{Planet c}  & \multicolumn{4}{c}{current age} & \multicolumn{4}{c}{at 3\,Myr} \rule{0pt}{2.3ex} \rule[-1ex]{0pt}{0pt}\\
    \hline              
%   2.6  & 11.6  & 11.8 & 3.5 & 1.2 & $2.2 \times 10^{11}$  & 15.3 & 11.1 & 24.2 & $8.6 \times 10^{12}$ \\
     2.6  & 11.7  & 11.8 & 3.5 & 1.5 & $2.2 \times 10^{11}$  & 18.7 & 20.4  & 37.6 & $4.0 \times 10^{13}$ \rule{0pt}{2.3ex} \rule[-1ex]{0pt}{0pt}\\
%     1.2  & 1.7  & 1.7 & 1.2 & 0.0 & /  & 57.9 & 16.0 & 97.0 & $1.5 \times 10^{14}$ \\
                
    \hline                  
    \end{tabular}
\end{table*}

For the stellar X-ray emission at different ages, we adopted the analytic description by \cite{Penz08a} anchored to the current value of the X-ray luminosity, $L_{\star,\, X} = 10^{29}$ erg s$^{-1}$ in the band 5--100\,\AA, derived from the rotation--activity relationships by \cite{Pizz03}. 
To evaluate the stellar irradiation in the XUV band, we adopted the scaling law between EUV (100--920\,\AA) and X-ray (5--100\,\AA) luminosities derived by \cite{SF22}:
\begin{equation}
\log L_{\rm EUV} = (0.793 \pm 0.058) \log L_{\rm x} + (6.53 \pm 1.61)~,
\label{eq:euv}
\end{equation}
which is an updated version of the widely used relationship by \cite{SF11}. The predicted evolution of the X-ray and EUV stellar luminosities are shown in Fig.\ \ref{fig:evoltrack}.

To determine the evolution of the planetary radius associated with the variation of the planetary mass, we adopted the core--envelope structure model described in Appendix\,\ref{app:structure}. With this model, we incorporate a time dependence of the envelope size linked to the variation of the surface equilibrium temperature during the evolution.

We then proceeded with the time evolution of the photoevaporation. For each time step of the simulation, we compute the mass-loss rate and update $f_{\rm atm}$ and the planetary mass, obtaining a new value of $R_{\rm env}$ with the relation by \cite{LopFor14}. The latter quantity added to the core radius (assumed constant) provides the updated planetary radius. 

% At each time step, the total planetary mass is determined by the mass loss rate at the previous step. The updated value of $f_{\rm atm}$ allows us to adjust the total radius with the analytic relation by \cite{LopFor14}.

According to the aforementioned scenario and assuming a core that does not change in size or mass, we followed the planetary evolution back in time. We stopped our simulations in the past at 3\,Myr, the end time of planetary formation models, that is, when the circumstellar disc has already disappeared and each planet is in its final, stable orbit.
%For the future evolution, we let the system evolve from the current age (650\,Myr for TOI\,5398) until 5\,Gyr.
The results of the simulations, described in Sect. \ref{sec:photoevap}, are shown in Figures \ref{fig:evaporation_mass_radius} and \ref{fig:evaporation_atmo}, and in Table \ref{table:evpar}.

The mass-loss rate for planet b closely follows the broken power-law decrease in the XUV irradiation over time (Fig. \ref{fig:evoltrack}). The variation for planet c appears smoother and faster. The reason for the different behaviour is that the mean density of planet b changes little during the evolution, while it increases substantially for planet c. This also explains why the mass-loss rate of planet c becomes lower than that of planet b at an age of around $ 530$\,Myr.

We checked the deviations of the actual numerical results of the ATES code from the analytic approximation we employed. Our computed mass-loss rates are systematically lower by a factor of 1.5--2 than the more precise values. This difference is negligible for the evolution of planet b but may imply an underestimation of the initial mass and radius for planet c (by less than a factor of 2). The mass-loss rate could be overestimated by a similar factor if the EUV flux were predicted with the scaling law proposed by \cite{Jo2021} instead of the adopted Eq. \ref{eq:euv}.

We stress that for the backward-in-time modelling, we are considering as initial conditions the current planetary core--envelope structure (Appendix\,\ref{app:structure}).
%, and we assume that the core mass and radius do not change in time. 
These initial conditions provide strong constraints on the possible evolutionary histories of the two planets.

\section{Dynamics of the system}
\label{app:dynamics}
The system's dynamics is currently dominated by the mutual secular perturbations of planets and tides. In the long term, the system is stable even if the outer planet (b) were closer than the nominal orbit (up to a limit of $a_b > 0.07$ AU) and a high eccentricity (0.2) is assumed. At present, the two planets are very close to a 9:4 mean motion resonance, which may have been broken after the dissipation of the gaseous disc. It is a weak resonance, but it may be responsible for an eccentricity value slightly different from zero for both planets, built during the resonance capture induced by the migration of the planets while embedded in the disc.

If planet b were in an almost circular orbit, the secular perturbations on planet c would cause only a low forced eccentricity (of the order of 0.001), which might trigger some tidal deformation --- as in the case of Io and Jupiter. On the other hand, if planet b were on an eccentric orbit with a value equal to the upper boundary (0.12) given by the data analysis (see Table \ref{table:model-rm}) then the forced eccentricity of planet c would be about 0.08 with a sinusoidal oscillation amplitude of 0.16. In the latter case, planet c would be likely to undergo strong tidal deformations, which would possibly represent an additional heat source.

%TC:endignore
\end{appendix}

\end{document}